\documentclass[%
 reprint,
 amsmath,amssymb,
 aps,
]{revtex4-1}
\usepackage{setspace,gensymb}
\usepackage{graphicx,svg}% Include Fig.files
\usepackage{dcolumn}% Align table columns on decimal point
\usepackage{bm}% bold math
\usepackage{siunitx}
\usepackage{hyperref}
\usepackage{physics}

\usepackage[symbol]{footmisc}

\hypersetup{
    colorlinks   = true, 
    urlcolor     = blue,
    linkcolor    = blue, 
    citecolor   = blue 
}

\begin{document}

%\preprint{APS/123-QED}

\title{Gate Error Analysis of Tunable Coupling Architecture\\ in the Large-scale Superconducting Quantum System}% Force line breaks with \\

\author{Dowon Baek,$^1$ Seong Hyeon Park,$^2$ Suhwan Choi,$^1$ Chanwoo Yoo,$^1$}%
\author{Seungyong Hahn$^2$}
\email[]{hahnsy@snu.ac.kr}

\affiliation{%
 $^1$Department of Physics and Astronomy, Seoul National University, Seoul 08826, Korea \\
 $^2$Department of Electrical and Computer Engineering, Seoul National University, Seoul 08826, Korea
}%

\date{December 8, 2022}

\begin{abstract}
In this paper, we examine various software and hardware strategies for implementing high-fidelity controlled--Z gate in the large-scale quantum system by solving the system's Hamiltonian with the Lindblad master equation. First, we show that the optimal single-parameter pulse achieved the gate error on the order of $10^{-4}$ for the 40~ns controlled--Z gate in the 4-qubit system. Second, we illustrate that the pulse optimized in the isolated 2-qubit system must be further optimized in the larger-scale system to achieve errors lower than the fault-tolerant threshold. Lastly, we explain that the hardware parameter regions with low gate fidelities are characterized by resonances in the large-scale quantum system. Our study provides software-oriented and hardware-level guidelines for building a large-scale fault-tolerant quantum system.
 \end{abstract}

\pacs{Valid PACS appear here}% PACS, the Physics and Astronomy
\maketitle

\section{\label{sec:level1}Introduction}
Superconducting quantum circuits have been widely studied as a promising platform for realizing fault-tolerant quantum computers, owing to their high controllability of qubits using microwave electronics \cite{kjaergaard2020superconducting}. In particular, the recent manifestation of quantum supremacy with intermediate-scale quantum devices \cite{arute2019quantum} has ignited the need to study high-fidelity and scalable quantum processors. To better understand the behavior of large-scale quantum processors, it is essential to consider quantum operations in the multi-qubit system, considering the leakage to spectator elements.

In the large-scale quantum system, it is commonly believed that quantum error-correction is indispensable for reliable quantum operations, and the target gate fidelities are set by error tolerance of the applicable error-correction scheme.  Quantum error-correction schemes that are widely studied in the literature include surface code \cite{andersen2020repeated,ai2021exponential,erhard2021entangling,marques2022logical}, bosonic code \cite{hu2019quantum,cai2021bosonic,gertler2021protecting}, and color code \cite{nigg2014quantum,chamberland2020triangular,reichardt2020fault}. Among these, surface code is the error-correction scheme native to the superconducting qubit platform as it only utilizes the nearest neighbor coupling \cite{barends2014superconducting}, which is naturally embedded in the planar device-based quantum computing platforms. Moreover, its low fidelity threshold of about $99\%$ for fault-tolerant error correction \cite{fowler2012surface,wang2011surface} makes the superconducting qubit an attractive platform for scaling up the quantum system. Recent realization of distance-three surface code on superconducting qubits \cite{zhao2022realization} have made this direction even more promising. Although the state-of-the-art experimentally demonstrated 2-qubit gate fidelities are currently reaching 99.9\% \cite{xu2020high,foxen2020demonstrating,negirneac2021high,sung2021realization}, the demonstrations are generally limited to the isolated 2-qubit system. Therefore, one of our important goals is to achieve a similar level of gate fidelities even in the large-scale quantum system.

Regarding the superconducting qubit architectures, fixed-frequency architecture \cite{chow2011simple,kim2022high,morvan2022optimizing}, and the tunable architecture \cite{bialczak2011fast,sung2021realization,li2020tunable,sete2021floating,stehlik2021tunable} are the two most common architectures used in the community. In the fixed-frequency architecture, all device parameters must be carefully set and fixed before fabrication to meet the design requirements and minimize unwanted interactions. It also accompanies unpredictable manufacturing errors, which is unfavorable to building large-scale high-fidelity quantum devices. The tunable architecture, however, employs \emph{tunable coupler} to mediate and control the coupling between qubits. Hence, it is possible to tune the interactions between qubits to the desired value even after fabrication. Although this benefit comes at the cost of hardware overhead and flux noise associated with tunable qubits, its ability to dynamically tune the interactions make the tunable architecture favorable for building large-scale quantum system.

With this background, this paper presents the first gate error analysis of tunable coupling architecture in a large-scale quantum system. Specifically, we examine the gate fidelity of controlled--Z (CZ) operations in the one-dimensional 4-qubit chain and discuss the software-oriented and hardware-level optimization methods. In addition to few studies on analyzing 2-qubit operations in the multi-qubit system \cite{zhao2022quantum,chu2021coupler}, we aim to suggest the guideline on hardware design and pulse scheduling to achieve high-fidelity quantum operations in the large-scale quantum system.

In section \ref{sec:s2}, we present the principle of CZ gate operations and the simulation model describing the large-scale quantum system. In section \ref{sec:s3}, we elucidate the software and hardware approach to reducing gate errors in the multi-qubit system. Section \ref{sec:s4} concludes this paper by discussing outlooks.

\section{\label{sec:s2}Model}
\begin{figure}[tb]
    \includegraphics[width=\linewidth]{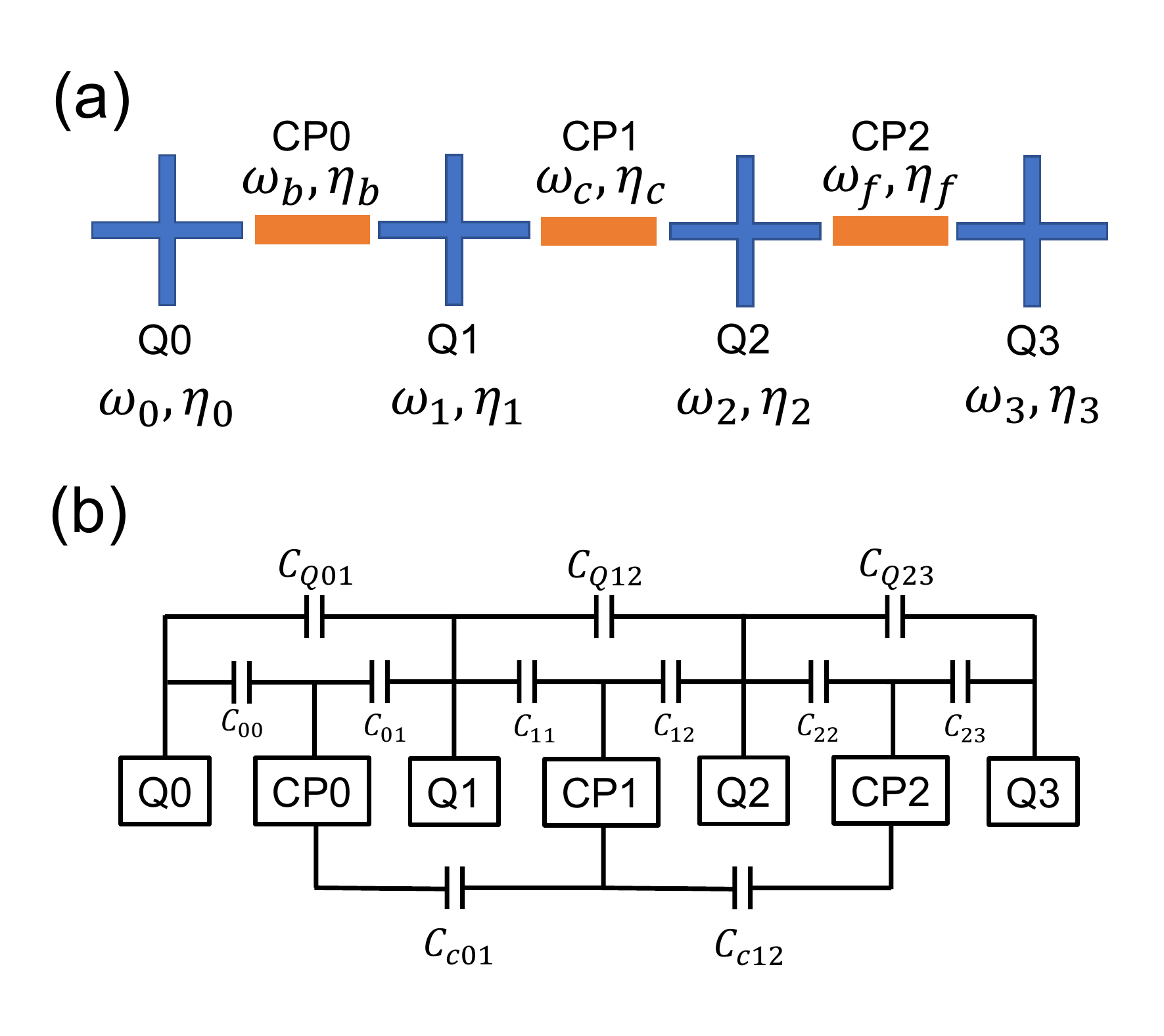}
    \caption{(a) Schematic of the 4-qubit superconducting quantum system used in the simulation. Q0-Q3 denote four Xmon qubits, and CP0-CP2 denote three tunable couplers. $\omega$ and $\eta$ below each element represent the resonant frequencies, and anharmonicities, respectively. The system parameters used in the simulation are presented in Table \ref{table:1}. (b) Circuit diagram of the 4-qubit system under consideration. Each qubit and coupler is a tunable element that consists of two Josephson junctions and a capacitor connected in parallel. Since the inter-element coupling decreases exponentially with the distance, we only consider NN and NNN coupling.}
    \label{fig1}
\end{figure}

We simulate the 4-qubit system, shown in Fig.\ref{fig1}(a), by solving the Lindblad master equation with QuTiP \cite{johansson2012qutip}. The three lowest-lying energy levels of each element are considered for the simulation. Following the standard circuit quantization procedure \cite{yan2018tunable}, the system Hamiltonian could be written as 
\begin{equation}
\label{eq1}
\begin{split}
    H=\sum_i \left(\omega_i a_i^\dagger a_i + \frac{\eta_i}{2}a_i^\dagger a_i^\dagger a_i a_i\right) \\
    - \sum_{i<j}g_{ij} (a_i-a_i^\dagger)(a_j-a_j^\dagger),
\end{split}
\end{equation}
where $a_i$ and $a_i^\dagger$ are annihilation and creation operators, respectively, and $\omega_i$, $\eta_i$, $g_{ij}$ denote resonant frequency, anharmonicity, and coupling constant, respectively. Here, the coupling constant $g_{ij}$ has the form $g_{ij}=r_{ij}\sqrt{\omega_i \omega_j}$, where $r_{ij}$ is the proportionality constant that can be expressed with capacitance matrix elements. Although this constant can be obtained from the finite-element simulation of the device, we set these to be fixed with the typical values: $r_{ij}=0.02$ for the nearest-neighbor (NN) pair, and $r_{ij}=0.0016$ for the next-nearest-neighbor (NNN) pair. The coupling between the pair that is farther than NNN is assumed to be negligible.

Next, we briefly explain how the CZ gate is performed in this system. CZ gate is the operation in which the Z-gate is applied to the target qubit if and only if the control qubit is in state $\ket{1}$. Equivalently, the CZ gate flips the phase of the 2-qubit system only when the system is in state $\ket{11}$, and such operation is achieved through the accumulation of tunable ZZ interaction between two coupled qubits.

\begin{table}[tb]
\centering
\begin{tabular}{c c c  c } 
 \hline\hline
 Variable & Value (GHz) & Variable & Value (GHz) \\ [0.5ex] 
 \hline\hline
 $\omega_0/2\pi$ & 5.05 & $\eta_0/2\pi$ & $-$0.3 \\ 
 $\omega_1/2\pi$ & 5.7 & $\eta_1/2\pi$ & $-$0.3 \\
 $\omega_2/2\pi$ & 5.0 & $\eta_2/2\pi$ & $-$0.3 \\
 $\omega_3/2\pi$ & 5.62 & $\eta_3/2\pi$ & $-$0.3\\ 
 $\omega_{b,idle}/2\pi$ & 7.83 & $\eta_b/2\pi$ & $-$0.25\\ 
 $\omega_{c,idle}/2\pi$ & 7.86 &$\eta_c/2\pi$ & $-$0.25\\ 
 $\omega_{f,idle}/2\pi$ & 7.70 & $\eta_f/2\pi$ & $-$0.25\\ [1ex]
 \hline \hline
\end{tabular}
\caption{System parameters used in the simulation. The idle frequencies of couplers are determined by minimizing the ZZ crosstalk between the two qubits. The subscript $idle$ denotes its value in the idle state.}
\label{table:1}
\end{table}

Since the Hamiltonian in Eq.\eqref{eq1} preserves the total number of excitations, we can consider the entire state space as a direct sum of manifolds, which are divided by the total number of excitations. In each manifold, the coupling between qubits will result in an avoided crossing. Particularly, in the double-excitation manifold, the state $\ket{11}$ is coupled to $\ket{20}$ and $\ket{02}$, and consequently, the energy level of $\ket{11}$ will deviate significantly from its bare energy level. Here, $\ket{ab}$ denotes the state in which the control qubit is in state $\ket{a}$ and the target qubit is in state $\ket{b}$. This results in the finite ZZ interaction \cite{zhao2020high},
\begin{equation}
    \nu_{ZZ} = (E_{\ket{11}}-E_{\ket{01}}) - (E_{\ket{10}}-E_{\ket{00}}),
\end{equation}
where $E_i$ denotes the energy level of state $i$. The phase flip of $\ket{11}$ is achieved through the accumulation of this ZZ interaction over the gate time $T_g$, \begin{equation}\int_0^{T_g} \nu_{ZZ} dt = \pi.\end{equation}

Particularly in the tunable coupling architecture, such ZZ interaction can be controlled by tuning the frequency of the tunable coupler. In our study, we compare the performance of four single-parameter pulses. There are two major reasons why single-parameterized pulses are desirable. Firstly, the search space during optimization is much smaller, and thus the optimization becomes much faster compared to multi-parameterized pulses. Moreover, adding more parameters typically provides only incremental improvement in gate performance compared to the effort required to find the optimum. Hence, we benchmark the performance of four single-parameter pulses: single-component Fourier pulse, quadratic pulse, hyperbolic pulse, and adiabatic pulse. The single-component Fourier pulse is given by,
\begin{equation}
    \frac{d\omega_c}{dt} = \sum_{k=1}^{k_{\textrm{max}}}\lambda_{Fk} \textrm{sin}\left(\frac{2\pi kt}{T_g}\right),
\end{equation}
where $\omega_c$ is the coupler's frequency, $T_g$ is the gate length, $k_{\textrm{max}}=1$, and $\lambda_{F1}$ is the single parameter to be optimized. Quadratic pulse is defined as
\begin{equation}
    \omega_c(t) = \omega_{c,idle} + \lambda_Q  t  (t-T_g),
\end{equation}
where $\omega_c(t)$ is the frequency of the coupler at time $t$, $\omega_{c,idle}$ is the coupler's idle frequency, and $\lambda_Q$ is the single parameter to be optimized.

Hyperbolic pulse is defined as 
\begin{equation}
    \omega_c(t) = \omega_{c,idle} + \textrm{cosh}(\lambda_H T_g /2)-\textrm{cosh}(\lambda_H(t-T_g/2)),
\end{equation}
where $\textrm{cosh}(x)=(e^x+e^{-x})/2$, and $\lambda_H$ is the single parameter to be optimized.
Lastly, the adiabatic pulse is defined as
\begin{equation}
    \frac{d\omega_c}{dt} = \frac{1}{G(\omega_c)}\lambda_A \textrm{sin}\left(\frac{2\pi t}{T_g}\right),
\end{equation}
where $\lambda_A$ is the single parameter to be optimized. $G(\omega_c)$ is the pre-factor defined as
\begin{equation}
    G(\omega_c)=\sum_u \sum_{v \neq u} \left| \frac{\langle u |\dot{v} \rangle}{\omega_u-\omega_v} \right|,
\end{equation}
where $\omega_c$ is the coupler's frequency, $\dot{v}=\partial/\partial\omega_c \ket{v}$, $\omega_u$ is the energy of the eigenstate $u$, the sum over $u$ runs over the truncated 2-qubit subspace, and the sum over $v$ runs over all eigenstates of the system \cite{chu2021coupler}. This pre-factor is a measure of the state's diabaticity. Note that the non-adiabatic transition to other eigenstates contributes to gate errors. The single parameter that parameterizes each pulse shape was optimized using the differential evolution method \cite{storn1997differential} by minimizing the CZ gate error

\begin{equation}
    1-F_{CZ} = 1- \frac{\left|\textrm{Tr}(U_{t}^{\dagger} U) \right|+\left|\textrm{Tr}(U_{t}^{\dagger} U) \right|^2}{d(d+1)},
\end{equation}
where $F_{CZ}$ is the CZ gate fidelity, $U_{t}$ is the ideal target CZ unitary, $U$ is the simulated unitary matrix truncated to the 2-qubit subspace, and $d$ is the system's dimension \cite{pedersen2007fidelity}. $U$ also includes single-qubit phase corrections, which are manually calibrated in real experiments.

In the next section, we first compare the CZ gate error of four single-parameter pulses introduced in this section. Secondly, using the best-performing pulse, we examine how the hardware design parameters, such as qubit frequency arrangements and qubits' anharmonicities, affect the gate error in the large-scale quantum system. Lastly, we discuss the impact of relaxation and decoherence times on the CZ gate fidelity.

\section{\label{sec:s3}Results \& Discussions}

\begin{figure}[tb]
    \includegraphics[width=.9\linewidth]{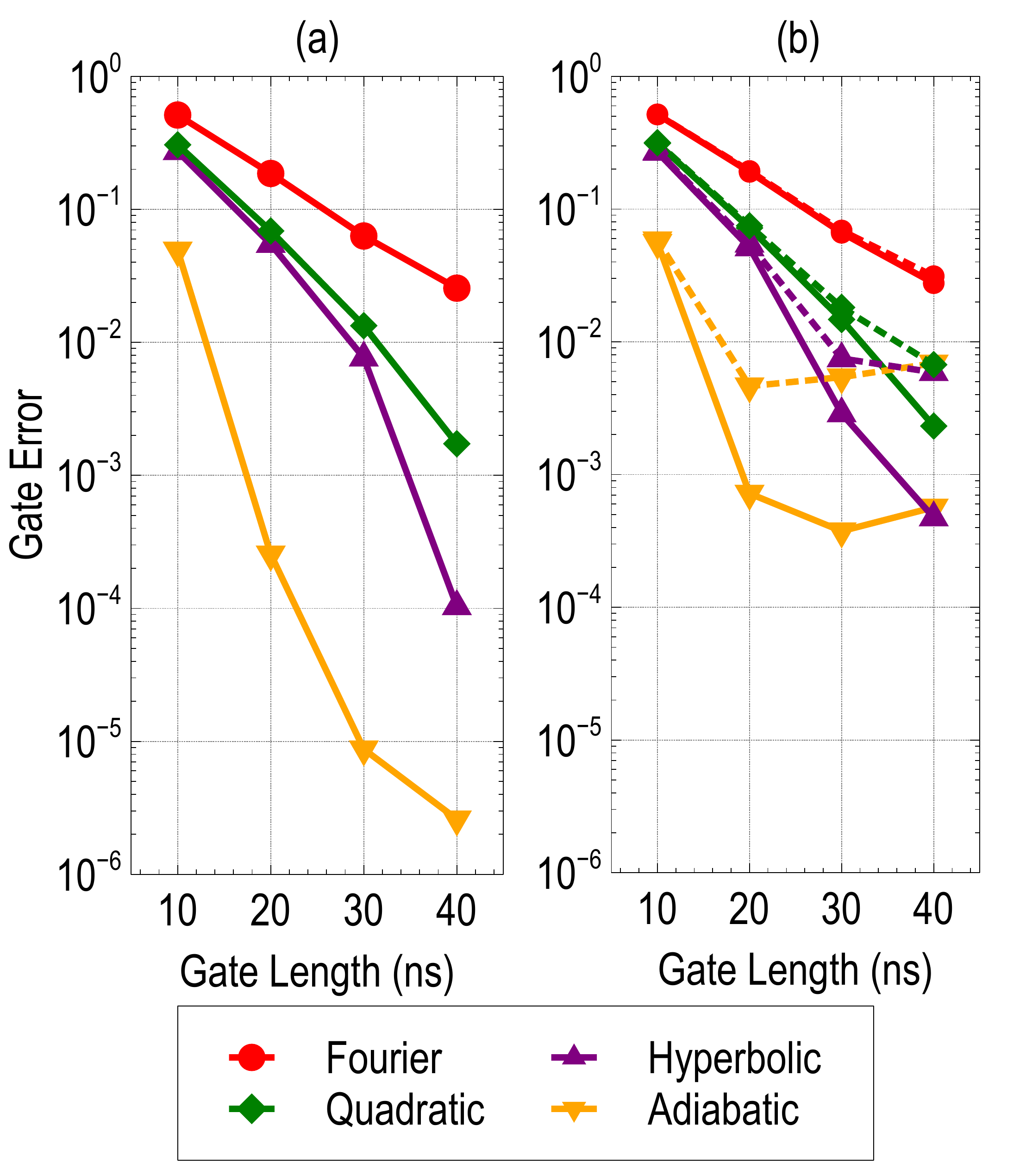}
    \caption{Comparison of CZ gate fidelities for various pulse shapes in (a) the 2-qubit and (b) the 4-qubit systems. In (b), dashed lines represent the gate errors calculated with the pulse optimized in the 2-qubit system, and solid lines represent the gate errors calculated with the pulse optimized in the 4-qubit system. System parameters in Table \ref{table:1} was used.}
    \label{fig2}
\end{figure}

The comparison of CZ gate errors between these four single-parameter pulses is shown in Fig.\ref{fig2}. Overall, we observe the trade-off between gate length and gate error; implementing fast, shorter gates inevitably accompanies more coherent errors. Among the four pulse shapes, the adiabatic pulse achieved significantly low gate error compared to other single-parameter pulses, with the gate error on the order of $10^{-6}$ for $40$~ns CZ gate in the 2-qubit system. In addition, the lowest achievable gate errors were much larger in the 4-qubit system than in the 2-qubit system. Such discrepancies became larger with increasing gate lengths due to the accumulation of population leakage to spectator qubits over time. The degree of discrepancy ranged from nearly zero for the Fourier pulse to two orders of difference for the adiabatic pulse.

Moreover, evaluating the gate errors in the 4-qubit system with the pulse optimized in the 2-qubit system did not yield the optimal result. In Fig.\ref{fig2}(b), dashed lines illustrate the gate errors calculated with the pulse optimized in the 2-qubit system, and solid lines show the minimum error achievable in the 4-qubit system. The disparity between these two lines was particularly large for the adiabatic pulse, illustrating an order difference in gate error for gate lengths longer than 20~ns. Considering that the fault-tolerant fidelity threshold for superconducting qubits is about $10^{-2}$ \cite{fowler2012surface,wang2011surface}, additional optimization in the larger system is significant and necessary. Since incoherent errors would increase the gate error, it would be safe to reduce the coherent error to much smaller than $10^{-2}$, rather than just below the threshold.

\begin{figure}[tb]
    \includegraphics[width=\linewidth]{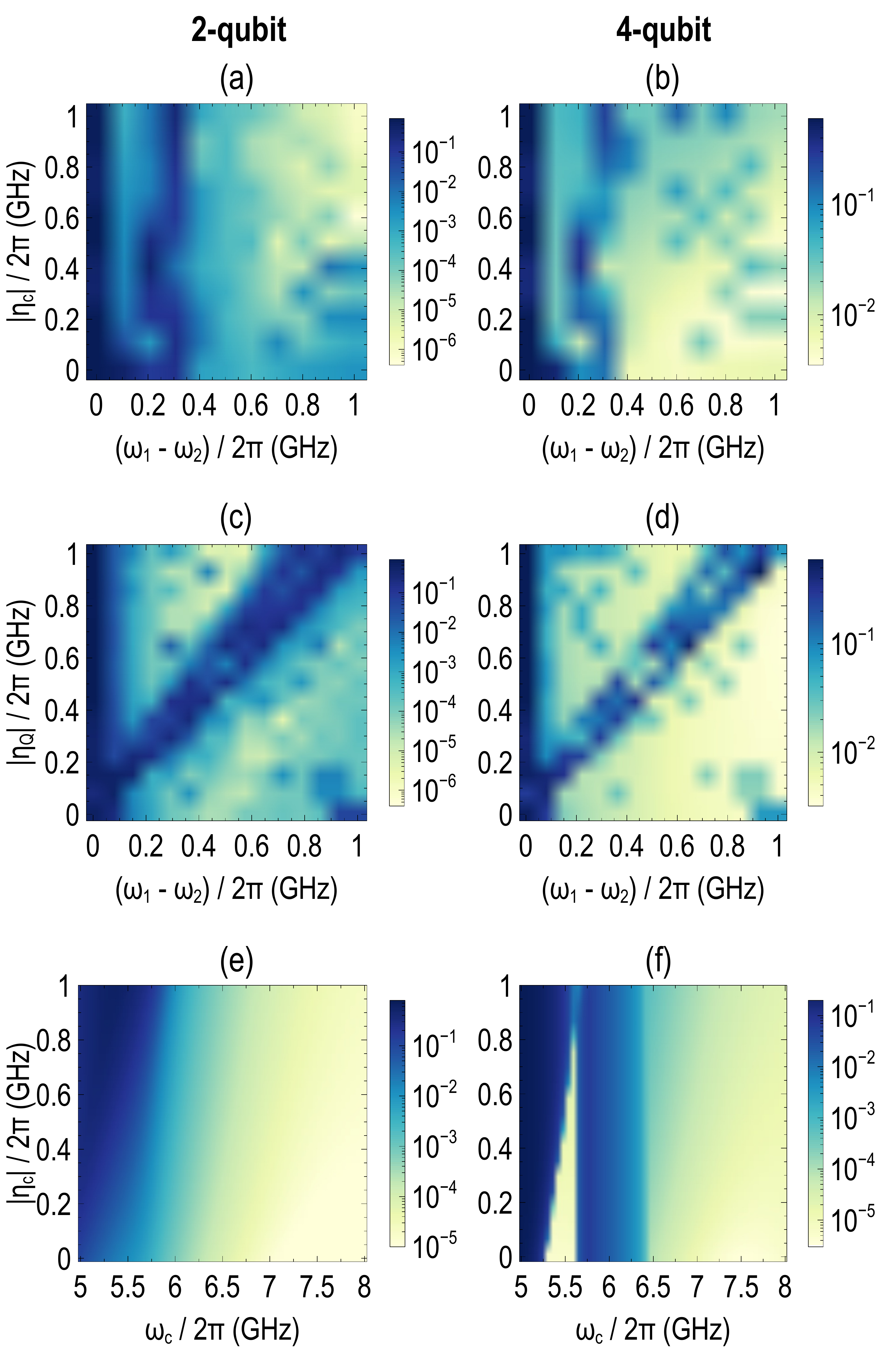}
    \caption{The plot of optimal gate errors for 25 ns CZ gate implemented with the adiabatic pulse as a function of coupler's anharmonicity $|\eta_c|$ and detuning between qubits $\Delta = (\omega_1-\omega_2)$ in (a) the 2-qubit, and (b) the 4-qubit systems. The plot of optimal gate errors for 25 ns CZ gate implemented with the adiabatic pulse as a function of qubits' anharmonicity $|\eta_Q|$ and $\Delta$ in (c) the 2-qubit, and (d) the 4-qubit systems. When changing the detuning between qubits, $\omega_2$ was fixed to 5.0~GHz, and $\omega_1$ was changed, where $\omega_1$ is the Q1's frequency, and $\omega_2$ is the Q2's frequency in Fig.\ref{fig1}. The frequencies of qubits Q0 and Q3 are fixed to values indicated in Table \ref{table:1}. The plot of the absolute value of ZZ interactions ($|\nu_{ZZ}|$) in GHz as a function of the coupler's $\omega_c$ and $|\eta_c|$ for (e) the 2-qubit, and (f) the 4-qubit systems.}
    \label{fig3}
\end{figure}

Since the adiabatic pulse achieved the best performance in both the 2-qubit and 4-qubit systems, we now consider the effect of hardware parameters on CZ gate errors with the adiabatic pulse. Fig.\ref{fig3}(a) and (c) show the optimized gate error as a function of detuning between qubits ($\Delta$) and qubits' and coupler's anharmonicities ($\eta_Q$, $\eta_c$) in the 2-qubit system. The gate errors became generally smaller with increasing detuning between qubits. This could be understood from smaller unwanted population exchange when the two qubits are detuned farther away from resonance. Particularly, the two resonance regions, each characterized by $\omega_1=\omega_2$ and $\omega_2=\omega_1+\eta_1$, showed significantly high errors. These lines are manifested as two parallel dark lines around $\Delta=0$ GHz and $\Delta=0.3$ GHz in Fig.\ref{fig3}(a), and the vertical line at $\Delta=0$ GHz and the diagonal dark line in Fig.\ref{fig3}(c). This could be attributed to large crosstalk between qubits on resonance. In the ideal CZ gate, we do not expect any population exchange between states. In addition, the gate errors were mostly oblivious to the qubits' and the coupler's anharmonicity except for the shift of resonance regions when the qubits' anharmonicities are changed.

 The errors evaluated in the 4-qubit system are shown in Fig.\ref{fig3}(b) and (d). As in the 2-qubit system, the gate errors are generally independent of the coupler's and qubits' anharmonicities. The gate error as a function of detuning between qubits exhibited its local maximum near resonances, each characterized by $\omega_1=\omega_2$ and $\omega_2=\omega_1+\eta_1$. The overall gate fidelities in the 4-qubit system were generally smaller than those in the 2-qubit system due to the intricate coupling between qubits, couplers, and spectator qubits.

 To better understand the performance degradation in the 4-qubit system, we compare the distribution of ZZ interactions between two qubits in the 2-qubit and 4-qubit systems. Since the CZ operation is performed through the accumulation of ZZ interactions ($\nu_{ZZ}$), the three conditions necessary for implementing high-fidelity CZ gate are the following: (1) there must exist a state in which $\nu_{ZZ}$ is highly suppressed, which is used for the idle state, (2) $\nu_{ZZ}$ should vary slowly since otherwise, non-adiabatic leakage to other states will lower the gate fidelity, and (3) there must exist a region with large enough $\nu_{ZZ}$ so that enough phase can be accumulated to perform the CZ gate. Among these, (1) and (2) can be readily achieved in the large-scale system by detuning the frequency of couplers far from the qubits, and the appropriate pulse scheduling, respectively. Hence, our utmost interest is on (3), or equivalently, to examine the factors that may potentially hinder phase accumulation in the large-scale system. We now compare the 2-qubit system and the 4-qubit system in this regard.

 The distribution of ZZ interactions as a function of the coupler's anharmonicity and the coupler's frequency is shown in Fig.\ref{fig3}(e) and (f). In Fig.\ref{fig3}(e), we observe the increase of ZZ interactions as the coupler's frequency becomes close to the qubits' frequencies. The strong ZZ interaction near the qubits' frequencies is what enables the CZ gate. Although a similar trend was observed in the 4-qubit system, Fig.\ref{fig3}(f) exhibits the seemingly discontinuous change of ZZ interaction around the coupler's frequency of 5.4~GHz. In fact, as opposed to monotonously increasing ZZ interaction with the decrease of the coupler's frequency in the 2-qubit system, the ZZ interaction in the 4-qubit system shows a rather complicated behavior, owing to the existence of spectator qubits and the coupling with their energy levels. One of such behavior includes the small negative ZZ interaction value near the qubits' frequencies. Recalling that the CZ gate is achieved through the accumulation of ZZ interactions, we conclude that this sign change of ZZ interaction may hinder the accumulation of phases, thereby compromising the gate fidelities. Overall, the complexity of energy levels in the 4-qubit system makes its lowest achievable gate fidelity much smaller than the 2-qubit system. It is preferable to set large detuning between qubits and stay far from resonances to achieve high gate fidelities.

\begin{figure}[tb]
    \includegraphics[width=\linewidth]{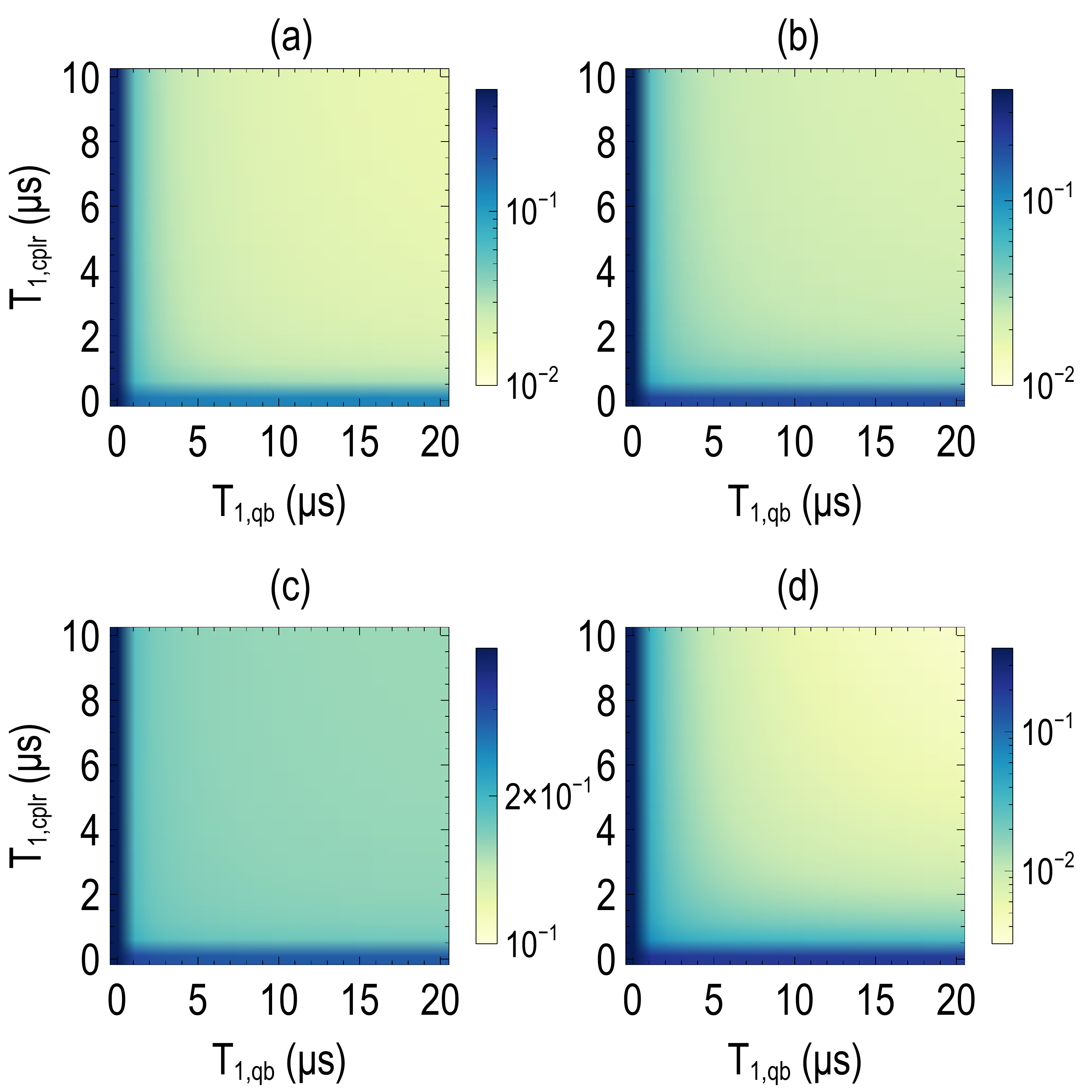}
    \caption{Loss of the population in the state $\ket{11}$, $p_{\ket{11}}$, at the end of the optimal 25~ns CZ gate as a function of qubits' relaxation time $T_{1,\textrm{qb}}$ and the coupler's relaxation time $T_{1,\textrm{cplr}}$ for (a) Fourier, (b) hyperbolic, (c) quadratic, and (d) adiabatic pulse in the 2-qubit system. The initial state was $\ket{11}$, and $T_2=T_1$ was assumed for both the qubits and the coupler.}
    \label{fig5}
\end{figure}

Additionally, it is also important to consider incoherent noise sources in the system. The most common noise sources in the superconducting qubit system include charge noise, flux noise, photon shot noise, and quasiparticle tunneling \cite{krantz2019quantum}. The effect of these incoherent errors can be abstracted through relaxation and decoherence times $T_1$ and $T_2$, which can be readily included in the Lindblad master equation as follows to evaluate their effects \cite{manzano2020short}:
\begin{equation}
    \dot{\rho}(t)=-\frac{i}{\hbar}[H(t),\rho(t)]+\frac{1}{T_1}\mathcal{L}[a]\rho(t)+\frac{1}{T_\phi}\mathcal{L}[a^\dagger a]\rho(t),
\end{equation}
where $\hbar$ is the reduced Planck's constant, $\rho(t)$ is the density matrix at time $t$, $1/T_\phi=1/T_2-1/2T_1$, and
\begin{equation}
\mathcal{L}[B]=B\rho(t)B^\dagger-\{B^\dagger B,\rho(t)\}/2,
\end{equation}
where $B$ is the input operator. For instance, Fig.\ref{fig5} shows the population loss of the initial state $\ket{11}$ at the end of the optimal 25~ns CZ gate as a function of qubits' and the coupler's relaxation times for four pulse shapes considered in this study. Overall, the population loss became smaller with the increasing relaxation times of qubits and the coupler. Particularly, the adiabatic pulse showed significantly smaller population loss in the relaxation-free limit since it is engineered to minimize non-adiabatic leakage. Although the effect of relaxations may not be apparent for short gate lengths, studying the material properties and qubit design with long relaxation and decoherence times is crucial to reduce multi-qubit gate errors in the large-scale quantum system.

\section{\label{sec:s4}CONCLUSION}
In this paper, we considered various software and hardware strategies for implementing high-fidelity CZ gates in the 4-qubit system by solving the system's Hamiltonian with the Lindblad master equation. First, we showed that the adiabatic pulse achieved the greatest performance, with an error on the order of $10^{-4}$ for the 40~ns CZ gate in the 4-qubit system, among the four single-parameter pulses considered in the study. Second, we illustrated that the pulse optimized in the isolated 2-qubit system must be further optimized in the larger-scale system to achieve errors lower than the fault-tolerant threshold. Third, we explained that the hardware parameter regions with low gate fidelities are characterized by resonances. Moreover, efforts to reduce coherence and relaxation times of qubits would be necessary to ensure high gate fidelities. Our study provides software-oriented and hardware-level guidelines for building a large-scale fault-tolerant quantum system.

\section*{Acknowledgement}
This research was supported in part by the SNU Student-Directed Education Undergraduate Research Program through the Faculty of Liberal Education, Seoul National University (2022). This research was also supported by National R\&D Program through the National Research Foundation of Korea (NRF) funded by Ministry of Science and ICT(2022M3I9A1072846), and by the Applied Superconductivity Center, Electric Power Research Institute of Seoul National University.

\bibliography{bibl}
\end{document}